\documentclass[aps,preprint,showpacs]{revtex4-2}
\usepackage{graphicx}

\begin{document}

\title{Spatial resolution enhancement in holographic imaging via angular spectrum expansion}

\author{Byung Gyu Chae}

\address{Holographic Contents Research Laboratory, Electronics and Telecommunications Research Institute, 218 Gajeong-ro, Yuseong-gu, 
Daejeon 34129, Republic of Korea}

\begin{abstract} 
Digital holography numerically restores three-dimensional image information using optically captured diffractive waves.
The required bandwidth is larger than that of hologram pixel at a closer distance in the Fresnel diffraction regime,
which results in the formation of aliased replica patterns in digital hologram.
From the analysis of sampling phenomenon,
the replica functions are revealed to be the components of higher angular spectra of hologram.
Undersampled hologram consists of the moire patterns formed by the modulation of original function by complex exponential function.
There is a one-to-one correspondence between the replicas in both real and Fourier spaces.
The possibility to acquire high-resolution images over a wide field view is explored in terms of the expansion process of angular spectrum by using replicas.
Only a low-NA hologram captured over a wide field restores a high-resolution image when using an optimization algorithm.
Numerical simulations and optical experiments are performed to investigate the proposed scheme.

\end{abstract}

\maketitle{}

\section{Introduction}
Optical microscopy is a fundamental tool to observe magnified images of a material through optical lenses. 
The spatial resolution of an image is limited by the wave characteristics of light, known as the Abbe diffraction limit,
which depends on the performance of the optical lens that collects the diffracted light at a higher angular scope \cite{1}.
Hence, an upper bound of the spatial resolution appears when an objective lens with a finite numerical aperture (NA) is used. 
To overcome the limited NA of a finite objective, various approaches have been successfully carried out [2-5],
where higher angular spectra of radiated light from materials are acquired in near-field and far-field imaging systems.
Recently, the necessity of microscopic imaging technology to observe samples in a wide range together with high precision has been indicated \cite{6,7,8,9}.
This technology would allow the sub-micrometric structure of materials, particularly those of pathological specimens in biomedicine, to be inspected effectively.

Digital holography is an interferometric imaging method in which the diffraction fringe of a diffracted wave is optically captured on an image sensor,
and the image is numerically reconstructed from the captured digital data based on scalar diffraction theory \cite{10,11,12,13}.
The angular spectrum of a digital hologram can be expanded by measuring multiple holograms using an oblique incident beam on the specimen,
thus resulting in an image resolution beyond the diffraction limit of existing optical systems.
Synthetic-aperture Fourier holographic microscopy is a representative tool for obtaining high-resolution images within a wide field-of-view \cite{3,6,7,14,15,16}.
Another diffraction imaging technology, i.e., Fourier ptychographic microscopy is used to extend the Fourier space \cite{17,18},
in which the complex amplitude is restored from the measured diffraction intensity via phase retrieval process. 
Although these imaging approaches effectively capture high-resolution image information in a wide field using only a low-NA objective,
there still remains a system complexity such as the multiple measurements or post data processing.

Recently, there have been several studies on high-NA digital hologram \cite{19,20,21}.
High-NA hologram suffers from the aliased replica fringe when captured at a closer distance.
In analysis of sampling theorem, the original image is well restored despite severe aliased errors. 
Because the aliasing errors arise from the required bandwidth larger than that of hologram pixel,
it is not certain whether or not there is a constraint on spatial frequency in the image restoration.

Holographic imaging is a representative linear system \cite{10}.
Thus, digital hologram of a finite-sized object can be analyzed by a point hologram.
Optical kernel functions directly correspond to elementary functions for point object in real and Fourier spaces.
When the functions are sampled in a digitized space, they can be represented as the summation of shifted form of original functions
because of complex amplitude with a quadratic phase term \cite{22,23}. 

In this study,
the relationship between aliased replica functions in both  real and Fourier spaces is mathematically or graphically investigated.
Firstly, it is found out that the replication patterns are moire interfernce fringes, i.e., the modulation of original function by complex exponential function,
which indicates that they are the components of higher angular spectra of hologram.
A one-to-one correspondence exists between them in both real and Fourier spaces.
A method for extending angular spectrum of hologram is studied using displacement of replication patterns in the space.
This technique enables high-resolution images to be restored, which reduces the complexity of imaging system. 
Numerical simulation is performed to investigate the proposed scheme, and the results of the optical experiment are presented.
Finally, the possibility of achieving the super-resolution beyond the Abbe diffraction limit using this strategy is discussed.

\section{Enhancement in spatial resolution of holographic image by angular spectrum expansion}
\subsection{Abbe theory of spatial resolution in digital holography}

Based on scalar diffraction theory \cite{10}, the diffractive wave field in the near-field region is characterized by optical kernel functions, i.e., an impulse response function in real space,
\begin{equation}
h(\rm{\bf{X}})= \it{} h_{0}  \exp \left(i \frac{\pi}{\lambda z} \rm{\bf{XX^T}} \right), 
\end{equation}
and a transfer function in the Fourier space,
\begin{equation}
H(\rm{\bf{U}})= \it{} H_{0}  \exp \left(-i \pi \lambda z \rm{\bf{UU^T}} \right), 
\end{equation}
where  $h_0 = \frac{e^{ikz}}{i \lambda z}$ and $H_0=e^{ikz}$. 
$\rm{\bf{X}}$ and $\rm{\bf{X^{'}}}$ represent two-dimensional (2D) vectors in $(x,y)$ and $(x',y')$ coordinates, respectively; and $k$ is the wavenumber of $ \frac{2\pi}{\lambda}$ with wavelength $\lambda$. 
The convolution notation, $g(\bf{X})= \it{O}(\bf{X})*\it{h}(\bf{X})$ describes the diffractive wave field $g(\bf{X})$ that propagates from the object field $O(\bf{X'})$. 
For a point object $\delta (\bf{X^{'}})$, the spherical wave diverges forward in a free space under the constraint of the angular spectrum, $k \geq 2\pi \sqrt{\bf{UU^T}}$, 
where $\bf{U}$=[$u$ $v$] is the spatial frequency vector.
The optical imaging performance depends on how much of higher components of angular spectrum in the diverging spherical wavelets is gathered through an optical lens.

Digital holography optically captures the wavefront of a diffracted wave from a material in digitized space, which is called the digital hologram. 
Figure 1(a) shows the geometric structure of a Fresnel holographic imaging system. 
The captured hologram data is band-limited in the spatial and frequency domains due to its finite aperture. 
Considering the spatial frequency vector of $\frac{\bf{X}}{\lambda z}$ in the Fresnel diffraction formula,
the space bandwidth $B_w$ in a digital hologram with $N$$\times$$N$ pixels of $\it{\Delta}$ pixel interval is represented as
\begin{equation}
B_w=\frac{N\it{\Delta}}{\lambda z}. 
\end{equation}
According to the Abbe diffraction theory, the bandwidth value determines the resolution limit $R_{\rm{lim}}$ of restored image \cite{21,24,25}:
\begin{equation}
R_{\rm{lim}}= \frac{\lambda z}{N\it{\Delta}}=\frac{\lambda}{2\rm{NA}}, 
\end{equation}
where $\rm{NA}=\it{n} \sin \theta=\frac{N\it{\Delta}}{2z}$, and $n$ is the refractive index of free space.
The hologram aperture function is an optical transfer function that defines the extent to which it collects the spatial frequency components of the diffracted wave.

\subsection{Replication properties of undersampled digital hologram in a digitized space}

When the hologram is acquired at a closer distance,
the bandwidth is larger than that of hologram pixel, which resluts in the aliasing errors in digital hologram.
In this case, it is important to establish that the image resolution would be finer than the hologram pixel value because of the restriction in spatial frequency, as depicted in Fig. 1. 

Holographic imaging is a linear system, and thus digital hologram of a finite object can be interpretable in terms of only an elementary function of point object \cite{10}.
The response function in real space becomes a hologram of point object from the relation, $g(\bf{X})= \delta (\bf{X})*\it{h}(\bf{X})$.

Let us investigate the sampling phenomenon of response function in digitized space.
The Fourier transform of the response function sampled by the sampling matrix, $\textbf{P}$ in real space is expressed as
\begin{equation}
\textbf{\emph{FT}} \left \{ \sum_{\bf{n}} h(\bf{nP}) \delta(\bf{X}-\bf{nP}) \right \} = \frac{\rm{1}}{|\rm{det} \, \bf{P}|} \sum_{\bf{m}} H (\bf{U}-\bf{mQ}),
\end{equation}
where $\textbf{\emph{FT}} \{ \}$ indicates the Fourier transform; $\textbf{Q}$ equals to $\textbf{P}^{-\rm{T}}$; and $\textbf{n}$ and $\textbf{m}$ are integers in 2D notation. 
Taking the inverse Fourier transform into the right-hand side of Eq. (5) yields the modulated form of the function, which can be rewritten in the shifted form of a quadratic phase functions \cite{22,23},
\begin{equation}
\sum_{\bf{n}} h(\bf{nP}) \delta(\bf{X}-\bf{nP}) = \frac{\rm{1}}{|\rm{det} \, \bf{P}|} \sum_{\bf{n}} \it{c_{\bf{n}} h}( \bf{X}-\lambda \it{z} \bf{nQ} ).
\end{equation}
The respective order is multiplied by the constant value, $c_n=e^{\frac{-i\pi\lambda zn^2}{\it{\Delta^{\rm{2}}}}}$.
The critical distance $z_c$ is defined when the repetition period $\frac{\lambda z}{\it{\Delta}}$ of function is the same as a field size $N\it{\Delta}$: 
\begin{equation}
z_c = \frac{N \it{\Delta}^2}{\lambda}.
\end{equation}
There appears no aliased fringe at above $z_c$. 

Similarly, when the transfer function is sampled, the same form as Eq. (6) can be obtained as follows:
\begin{equation}
\sum_{\bf{m}} H(\bf{mP_0}) \delta(\bf{U}-\bf{mP_0}) = \frac{\rm{1}}{|\rm{det} \, \bf{P_0}|} \sum_{\bf{m}} \it{c_{\bf{m}} H} \left( \bf{U}-\frac{mQ_0}{\lambda \it{z}} \right),
\end{equation}
where $\textbf{P}_{0}$ is the sampling matrix in the Fourier space and $\textbf{Q}_0=\textbf{P}_0^{-\rm{T}}$.
The transfer function is an angular spectrum of response function.

The undersampling properties of response function could be analyzed on the basis of Eq. (6). 
Principally, the sampling characteristics depend on the parameters $(z, \it{\Delta})$.
As the sampling interval $\it{\Delta}$ increases or the distance $z$ is smaller compared to a critical distance, the aliased replica patterns are generated.
The shifted quadratic phase function in Eq. (6) appears as the modulation of original function by complex exponential function:
\begin{equation}
h( \bf{X}-\lambda \it{z} \bf{nQ}) = \it{c_o h} ( \bf{X}) \exp \left(-\it{i} \pi \rm{\bf{X}} \bf{nQ} \right).
\end{equation}
The modulating term, i.e., complex exponential function is constant at specifications $(z_c, \it{\Delta})$,
where a shifted term becomes original function itself, and thus no information is extracted.

Figure 2 depicts the graphical analysis for the undersampled digital hologram of point object.
The point hologram in 2D space consisting of 256$\times$256 pixels with a 8-$\rm{\mu}$m pixel pitch was synthesized using a plane wave with a wavelength of 532 nm.
The hologram was made at a distance of a half of $z_c$, which forms four replica Fresnel zones.
In this case,
the complex exponential function in replica function, $e^{\frac{-i\pi \rm{\bf{X}}} {\it{\Delta}}}$ is the sinusoidal wave with a period of ${2\it{\Delta}}$, in Fig. 2(b).
The Fourier-transformed data is the shifted high-frequency component.  
All the shifted replica functions become the higher spatial-frequency components corresponding to those of original function.
The spatial frequency, $\frac{\rm{\bf{X}}} {\lambda z}$ of the concentric primary pattern linearly increases with a spatial distance.
It is intersesting that although the replica zones seem to be the moire fringes with a low frequency in appearance,
the aliased fringes should be higher spectrum components according to a spatial distance.
Spectrum distributions considering a finite pixel size are drawn briefly in Fig. 2(c).

Furthermore, a curve shape of replicas by all the sampling periods are embedded in the form of moire fringe in the primary function, as illustrated in Fig. 3.
The replica functions are formed at a reduced period of $\frac{\lambda z}{\it{s \Delta}}$ when the function is undersampled by $s$ multiples of $\it{\Delta}$.
On the other hand, the number of replicas, $N$ are larger by a demagnification process, i.e., zoom-out process.
This indicates that the self-similar patterns are formed by means of an extrapolation operation with a scale parameter of $\varepsilon$.
Theoretically, this phenomenon could be explained as a result of negative fractal dimension \cite{26}, $D = \lim_{\varepsilon \to 0} \frac{\log N(\varepsilon)}{\log(1/\varepsilon)} = -2$.

Figure 4 describes this phenomenon in the Wigner space.
The point hologram made at a distance of one-third of $z_c$ forms nine replica zones.
The Fresnel transform is represented as a shearing of the Wigner distribution function, $W(x,u)=W(x-\lambda zu,u)$. 
The overlapping of high-order spectra is inevitable at a distance below $z_c$, which causes aliasing fringes.
The Wigner distribution is rotated by $90^{\circ}$ via the Fourier transform \cite{27}, which also represents the replications of angular spectrum.
This shows clearly a one-to-one correspondance with both replica functions in real and Fourier spaces.
The aliased replica components of angular spectrum correspond to the replication elements of digital hologram.

\subsection{Method for spatial-resolution enhancement in image retrieval process}

The replica functions in the undersampled hologram play a role in higher angular spectrums of primary function.
In the image retrieval process,
the whole area of the digital hologram should be an aperture that satisfies the resolution relationship of Eq. (4). 
The hologram aperture extends to the whole area without a confinement of primary zone. 

Obviously, 
the hologram aperture size could be extended by spatially replicating the primary function once the primary function is known.
In this strategy, the primary functions both real and Fourier spaces can be applicable.
This expands the angular spectrum of hologram, and thus results in the retrieval of high-resolution image.
Mathematically, this concept would be possible because the replica function is the same as the original one except for the multiplication constant in Eqs. (6) and (8).
However, in the optically captured hologram, the addition of external noises due to a measurement environment is inevitable,
where it is hard to extract exact information about primary hologram function.

Recently, the research to expand the hologram aperture was performed by self-extraploation technique using a phase retrieval algorithm \cite{13}.
This method could be used for finding replica patterns from the primary function.
Here, the computation performance would be enhanced because the replica patterns have a self-similarity, 
and the computational capacity is reduced with comparison to the non-aliased data.

As another approach, the high-resolution image is retrieved from the undersampled hologram optically captured while keeping high aperture size.
In this case, the image can be effectively restored by replicating the angular spectrum in the Fourier space during the iterative optimization algorithm.
The iterative algorithm uses the angular spectrum method:
\begin{equation}
O(\textbf{X}') = \textbf{\emph{IFT}} \left[ \textbf{\emph{FT}} \{ \it{g} (\bf{X}) \} \times \it{H^{-\rm{1}}} (\textbf{U}) \right].
\end{equation}
Unlike the description of Fig. 1,
the pixels in both planes have the same value irrespective of the distance because of the sampling relation, $\it{\Delta}=\frac{\rm{1}}{N \it{\Delta_u}}=\it{\Delta'}$,
where an allowable image size is consistent with the hologram size. 
Undersampling by $m$-multiple pixels creates $m$-multiple replicas of the primary function.
The $m$-fold expansion of angular spectrum during the reconstruction process helps to restore the high-resolution image.

Finally,
the resolution performance would be enhanced when high-order diffractions from a material are used. 
High-order diffractions directly relate to the higher replica terms in Eq. (6).
Optically, these terms can be measured through a grating device located in front of sample. 
It was reported that the captured hologram to collect the highly diffracted beams on centered region improves the image resolution \cite{28}.
Here, there is no need for a beam collecting process.
Figure 5 is the schematic diagram for above proposed methods.

\section{Numerical analysis of resolution increment of holographic image}

Figure 6 illustrates the reconstruction properties of image by using an extraction process of replica patterns.
The initial complex-valued hologram for the object with 512$\times$512 pixels was synthesized by the Fresnel diffraction formula,
and then cropped by a factor of 2 to obtain the low-NA hologram with 256$\times$256 pixels, in Fig. 6(a).
The real-valued object consisting of several points with 2-$\mu$m pixel was used, 
and the initial hologram made at a half of $z_c$ has pixel resolution of  4 $\mu$m according to a sampling relation between hologram and object planes, $\it{\Delta'}=\frac{\lambda z}{N\it{\Delta}}$.

The replica patterns were extracted through a phase retrieval algorithm based on the Gerchberg-Saxton algorithm.
The absolute value of restored image was used during the repetition procedure, which is sufficient for a constraint. 
Original replication patterns are successfully found at a large repetition, which results in a retrieval of original image.
The one-dimensional profile of hologram in Fig. 6(e) clearly shows that the extrapolated data well matches with original one.
As not displayed here, it was confirmed that the hologram made at a fourth of $z_c$ retrieves original replication patterns. 

Figure 7 is the simulation results by using the undersampled hologram.
Initial high-NA hologram for the USAF resolution target was calculated by the inverse process of Eq. (10),
Both object and hologram spaces consist of 512$\times$512 pixels of a 4-$\mu$m pixel, and 
undersampled hologram prepared by 2-fold downsampling has 256$\times$256 pixels of an 8-$\mu$m pixel.
Undersampling is performed by periodically placing sub-pixels of the same size as the original pixels. 
This type of hologram can be optically measured using 2D grid mask attached to the image sensor. 
The hologram data acquired through the $m$-fold sub-pixel mask enables $m$-fold expansion of the angular spectrum.

Figure 7(a) is the restored image through the zero-padding technology, where the retrieval of initial resolution is unavailable.
However, as illustrated in Fig. 7(d), the image with the initial resolution of 4 $\mu$m is reconstructed from two-multiple expansion of the angular spectrum, 
which becomes a strong evidence that the replicated angular spectrum functions as the higher spectrum. Both magnified images show a clear discrimination. 
To remove the interference of high-order images, the extended angular spectrum was multiplied by a sinc function, $\rm{sinc} (\pi \it{\Delta} \textbf{U})$. 
There are still remnents of high-order noises and it shows relatively a poor image quality.

The iterative optimization algorithm based on the gradient descent method with a total variational regularization $\textbf{TV}$ was used to optimize the restored image \cite{29,30}:
\begin{equation}
\arg \min_{O} \frac{1}{2}  \parallel \textbf{\emph{Fr} \rm{(} \it{O} \rm{)} - \it{g}}  \parallel^2 + \gamma \textbf{TV}(O).
\end{equation}
The objective function consists of the least squares together with the $\textbf{TV}$ term, and $\gamma$ is a regularized parameter.
$\textbf{\emph{Fr}}$() denotes the Fresnel transform of Eq. (10).
The reconstructed image $O_k$ at respective iteration is updated by minimizing the loss function,
 i.e., the squared error between the original hologram $g_0$ and estimated one $g_k$.
The angular spectrum of the errors was expanded via its repetition process of $m$ multiples, resulting in considering the high frequency components.
The transfer function was also extended to $m$-multiple frequency space
and the estimated hologram was downsampled to match original resolution.
The convergence rate for optimization was arbitrarily controlled by changing the external parameters such as a learning rate $\alpha$ or regularization factor $\gamma$.

The iterative algorithm restores a high-quality image comparable to original image, when using the 4-fold downsampled hologram for the Shepp-Logan phantom, in Fig. 8.
The PSNR value of the reproduced image is estimated to be approximately 73 dB. 
As not shown here, we also confirmed that two-fold or eight-fold undersampled hologram well restores the image. 
The sinc function was used only in reconstruction process using the eight-fold undersampled hologram, and it was not necessarily required for this algorithm. 
The spatial resolution of the reconstructed image is recovered to the original value in accordance to the expansion extent of the angular spectrum. 
Figure 8(f) is the cross-sectional profile of the restored image after optimization. The initial image with a spatial resolution of 4 $\mu$m is exactly recoverable by using an iterative optimization algorithm.

\includegraphics[scale=1, trim= 2cm 18.5cm 1cm 0cm]{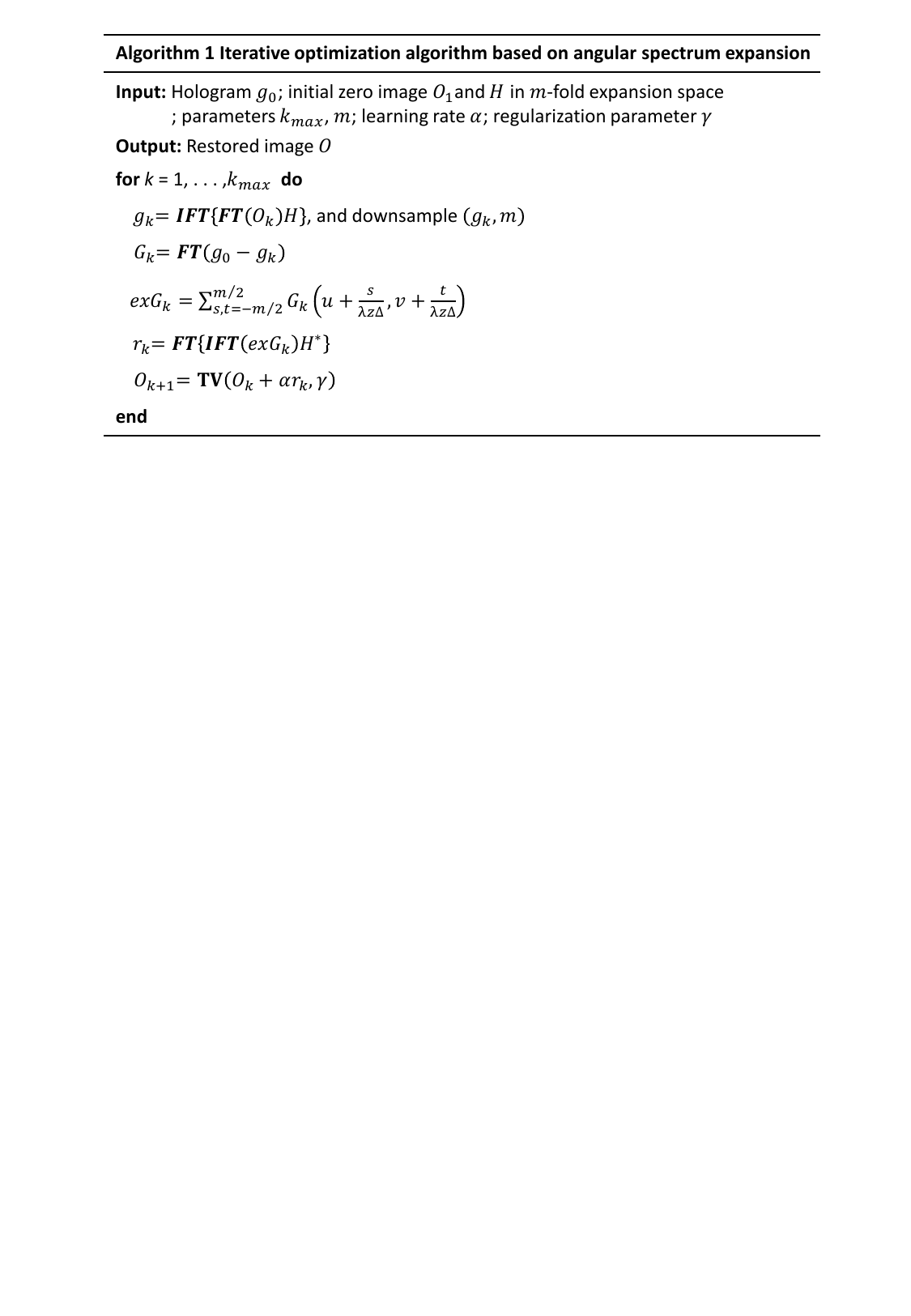}

\section{Optical experiments and discussion}

Figure 9(a) illustrates an off-axis holographic technique to acquire an optical hologram for the USAF resolution target.
A 532-nm green laser (LCX-532, Oxxius) was used as a light source.
A hologram with 2160$\times$2160 pixels was captured on an image sensor (GS3-U3-89S6M-C, FLIR) at a distance of 86 mm.
The captured hologram has a pixel pitch of 3.45 $\mu$m. 
The first-order region of the Fourier-transformed data in Fig. 9(c) was cropped to avoid non-diffraction beams and twin terms, and then moved to the center.
The inverse Fourier transform generated the final complex-amplitude hologram.

The reconstruction of the optical hologram is shown in Fig. 10. 
The low-resolution hologram was constructed via a 4-fold undersampling procedure, which is analogous to a hologram achieved using a 4-fold sub-pixel mask. 
The inferior image as compared with the original image was restored from the low-resolution hologram.
However, the expansion of angular spectrum in the iterative optimization algorithm recovers original resolution, as confirmed in the magnified pictures in Fig. 10(d).

A digital hologram is a captured diffraction wave. 
As mentioned previously, 
once the primary hologram function is known, a high-resolution image can be completely recoverable in the numerical reconstruction process.
It is desirable that the primary function is obtained using a sub-pixel mask, where the enhancement in the spatial resolution is shown to be limited by its undersampling extent. 
Principally, as referred to Fig. 3(a), this function exhibits a perfect form with no aliasing errors at a specified sampling rate. 
Therefore, the pixel value of whole area does not differ significantly from the quantity in the sub-pixel area, thus indicating that the primary function can be acquired without using a sub-pixel mask. 
In the numerical study of Figs. 7 and 8, we observed that even a low-resolution hologram prepared by an averaging operation of sub-pixel values clearly restores the image. 
On the other hand, to relaize this phenomenon in optical hologram, the precise experiment to capture the hologram well defining the NA should be implemented.

The imaging technology applied in this study overcomes the resolution limit of finite imaging systems, 
thus allowing high-resolution image to be restored over a wide field-of-view. 
The spatial resolution and field view of the measured image are determined based on the geometric structure of the imaging system. 
The use of an appropriate objective lens increases an imaging performance.

According to the Abbe resolution criterion, the maximum spatial resolution reaches a half of wavelength.
The optical transfer function used in the Fresnel diffraction regime can be extended to the form, 
$\exp (iz \sqrt{k^2-4\pi^2 \bf{UU^T}} )$ in the Rayleigh-Sommerfeld region. 
Although the extension of angular spectra through their replication is possible, the replica patterns in the high-NA appear to be distorted. 
Furthermore, it is not certain that this scheme is applicable to the evanescent field region.
When these two points are theoretically comprehensible, a super-resolution imaging using this strategy would be meaningful.

\section{Conclusions}
The replication patterns of the digital hologram and its Fourier-transformed form in both real and Fourier spaces correlate with higher angular spectra.
This phenomenon is found from the ananlysis of the sampling properties of hologram in a digitized space. 
The effective method for obtaining high-resolution images over a wide field view was developed to recover the complexity of the optical imaging system.
The hologram numerical aperture can be increased only by expanding the angular spectrum, resulting in the resolution enhancement of restored image.
Several methods of this holographic imaging were investigated through numerical simulations and optical experiments. 
An optical imaging system using a relatively low-NA objective lens can avoid disadvantages such as a narrow field-of-view and small depth-of-field.

This work was supported by Institute for Information \& Communications Technology Promotion (IITP) grant funded by the Korea government (MSIP) (2021-0-00745)

The authors declare no conflicts of interest.

\begin{figure}
\centering\includegraphics[scale=0.9, trim= 1cm 10.5cm 1cm 0cm]{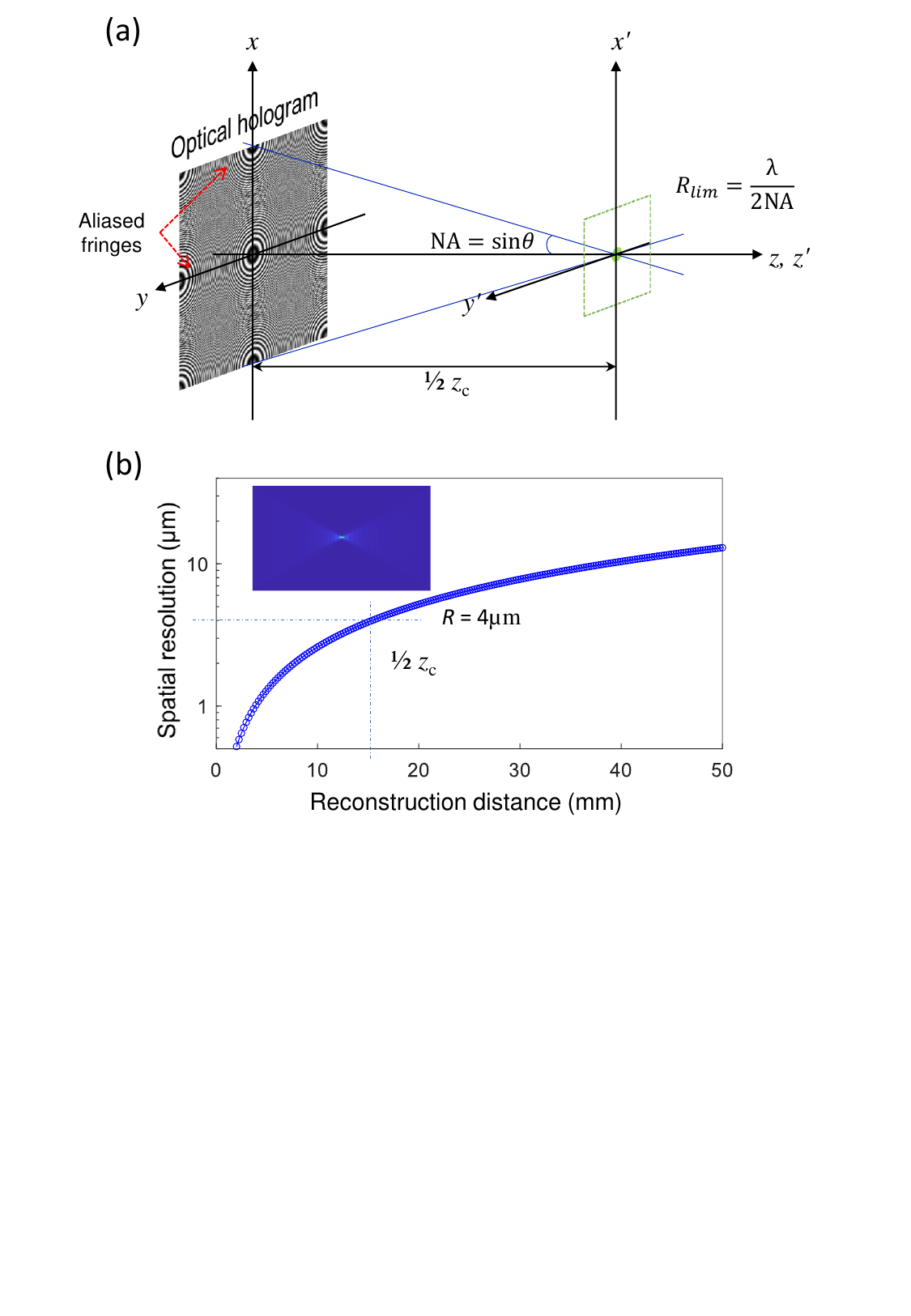}
\caption{(a) Schematic illustrations of the Fresnel holographic imaging system.
Point image is restored from digital hologram located at one-half of $z_c$.
Green box denotes allowable image space defined by diffraction of hologram pixel. 
(b) Change in spatial resolution as a function of reconstruction distance. 
Values for digital hologram consisting of 256$\times$256 pixels with a 8-$\mu$m pixel pitch are drawn.}
\end{figure}

\begin{figure}
\centering\includegraphics[scale=0.9, trim= 1cm 12cm 1cm 0cm]{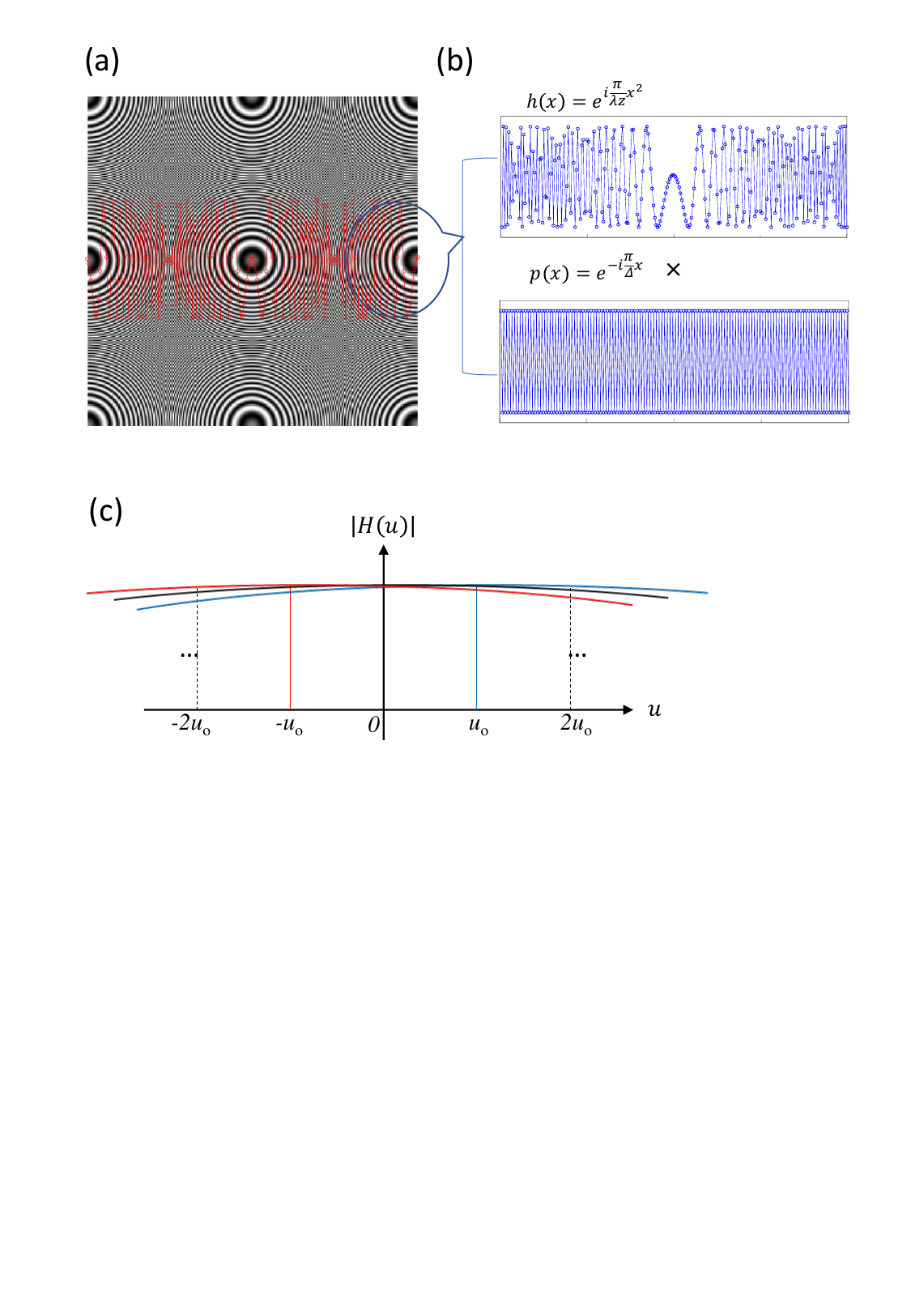}
\caption{Graphical analysis for the undersampled digital hologram of point object.
(a) Real-valued hologram synthesized at a distance of a half of $z_c$. One-dimensional quadratic phase function along the lateral line at the center is overlapped on the Fresnel zones.
(b) Original function and complex exponential function corresponding to $+1^{\rm{st}}$-order replica zone.
(c) Spectrum distributions considering a finite pixel size.}
\end{figure}

\begin{figure}
\centering\includegraphics[scale=0.9, trim= 1cm 14.5cm 1cm 0cm]{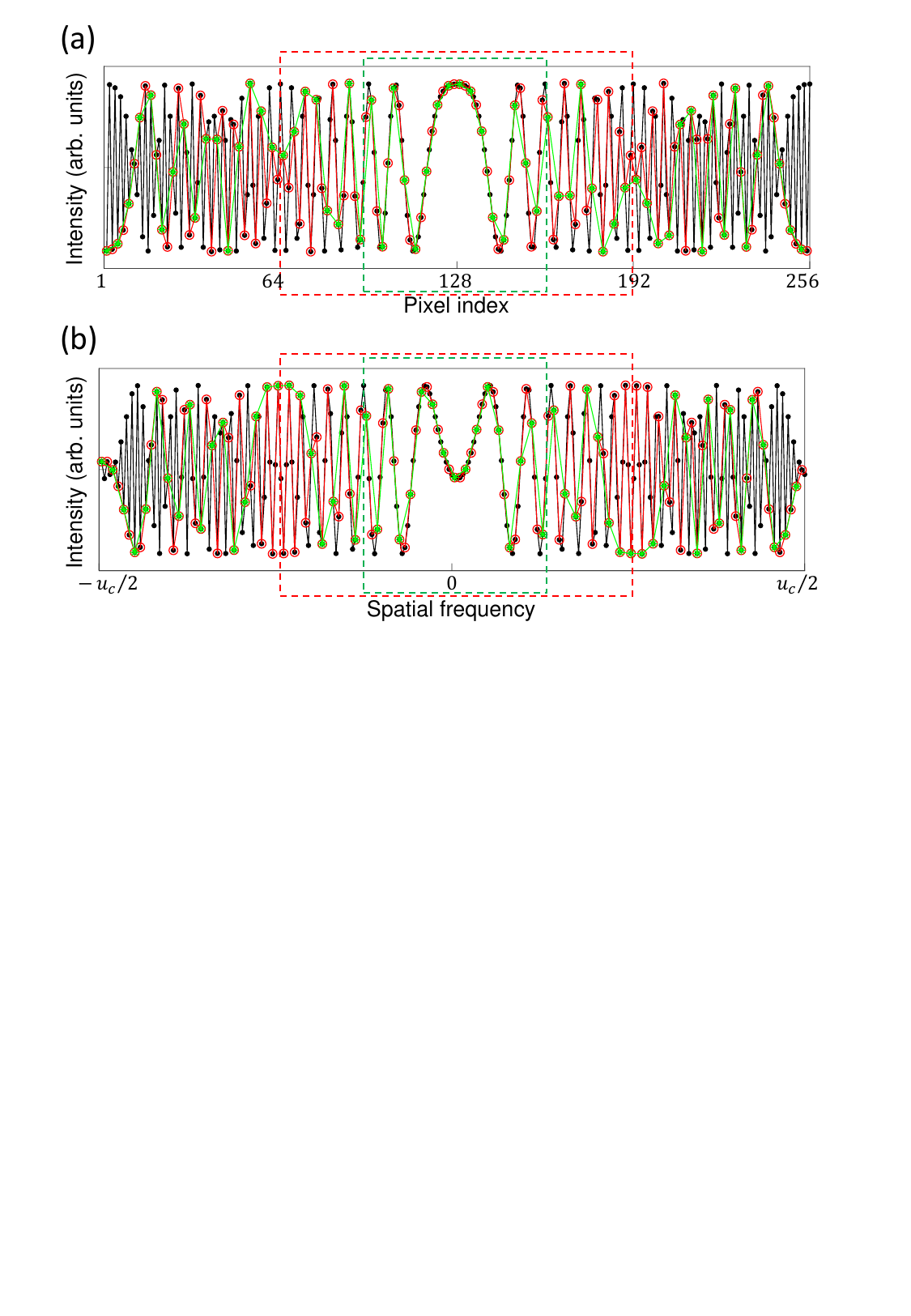}
\caption{Sampling phenomena of (a) response function and (b) transfer function of a point hologram in one-dimensional space.
Black, red, and green markers represent original transfer function, and undersampled functions by factors of 2 and 4, respectively.
Boxes indicate the corresponding primary functions.
$u_c$ is the maximum spatial frequency with respect to bandwidth of original function.}
\end{figure}

\begin{figure}
\centering\includegraphics[scale=0.9, trim= 2cm 11cm 1cm 0cm]{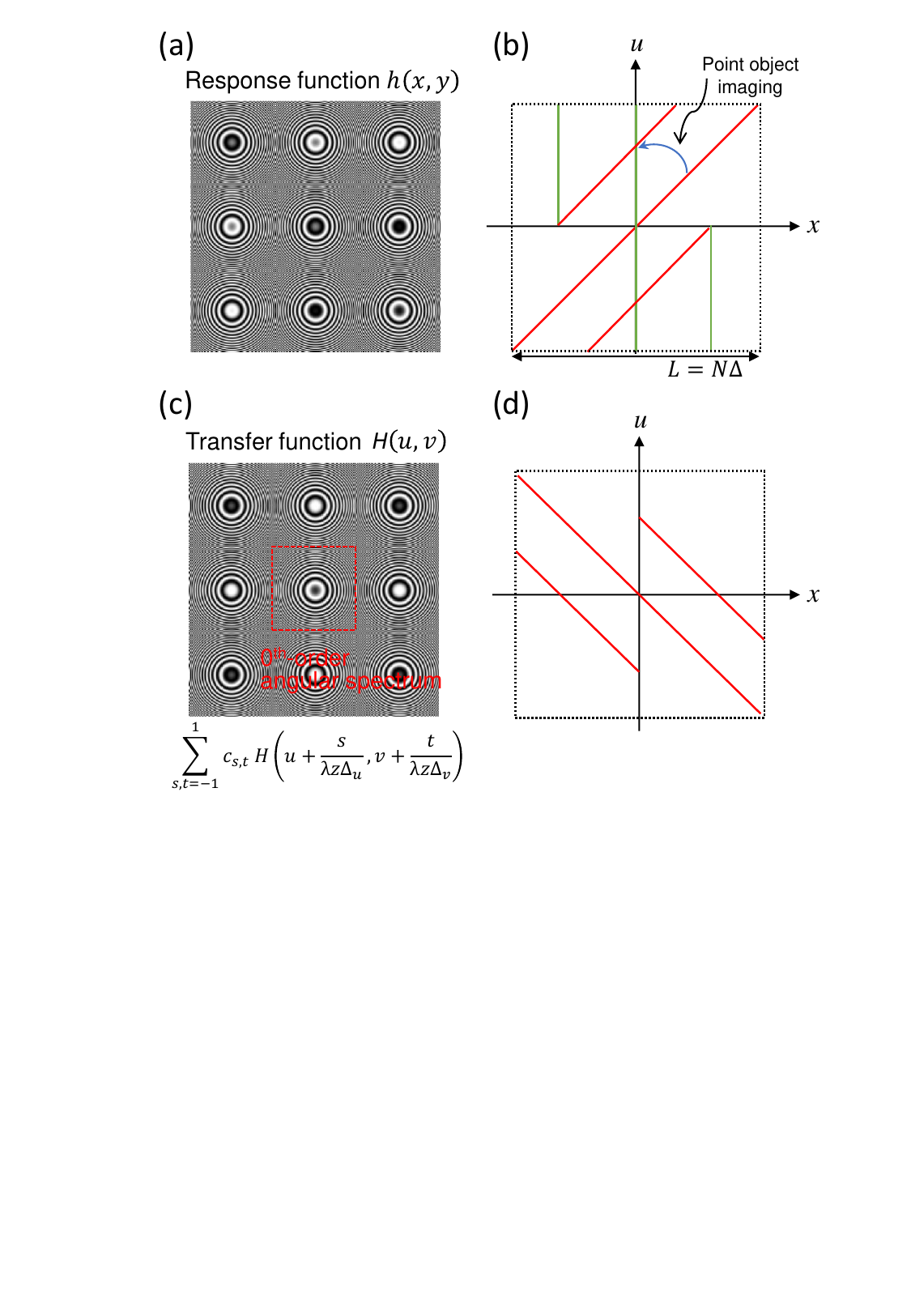}
\caption{Aliasing effects of undersampled hologram in the Wigner space.
(a) Digital hologram for a point object synthesized at a distance of one-third of $z_c$. (b) Wigner distribution description of digital hologram.
Point image is focused from inclined line designating spherical fringe of hologram.
(c) 3-fold expanded angular spectrum corresponding to digital hologram. (d) Wigner distribution description of angular spectrum.}
\end{figure}
 
\begin{figure}
\centering\includegraphics[scale=0.85, trim= 1cm 6cm 1cm 0cm]{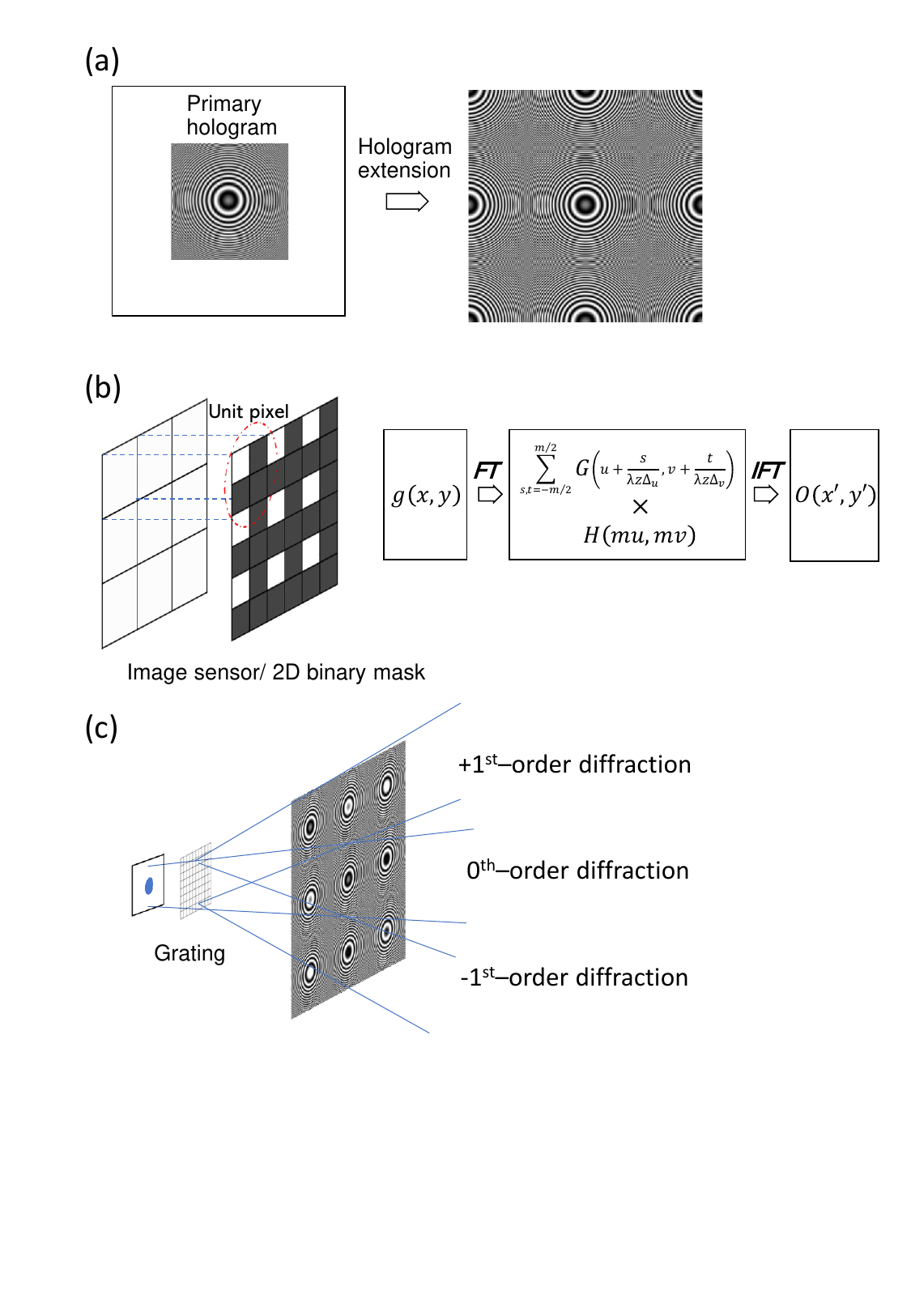}
\caption{Methods for increasing image resolution by expanding angular spectrum.
(a) Expansion of hologram aperture size by directly searching replica patterns.
(b) Image retrieval through the expansion process of angular spectrum for acquired low-resolution hologram.
(c) Image retrieval by using captured high-order diffractions.}
\end{figure}

\begin{figure}
\centering\includegraphics[scale=0.9, trim= 1cm 9.5cm 1cm 0cm]{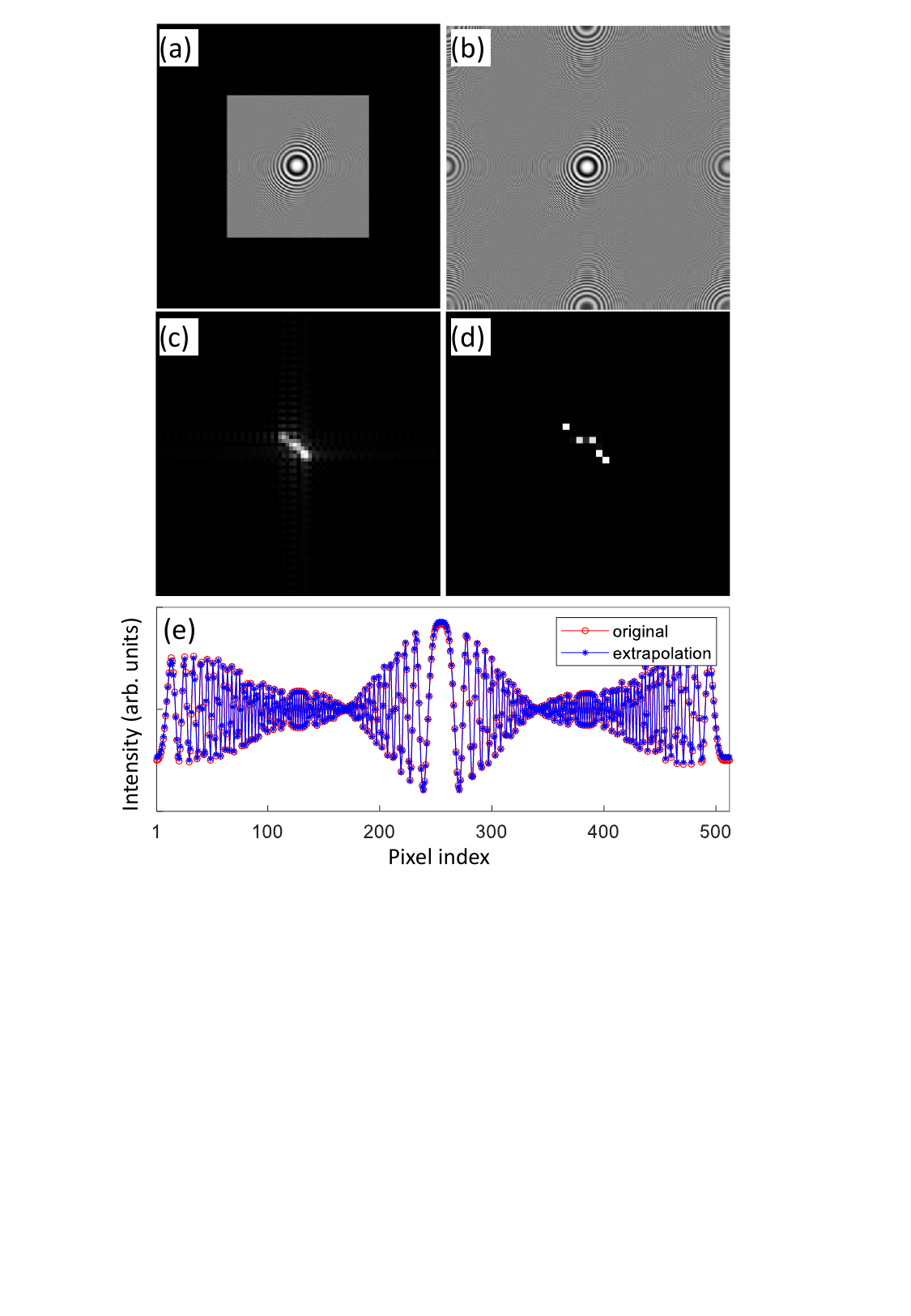}
\caption{Reconstruction properties of image by using an extraction process of replica patterns. 
(a) Low-NA hologram with 256$\times$256 pixels of an 8-$\mu$m pixel interval for several point objects. 
(b) Hologram extrapolated by using a phase retrieval algorithm. 
Reconstructed images for (c) low-NA hologram (d) extrapolated hologram.
(e) One-dimensional profiles of original and extrapolated holograms.}
\end{figure}

\begin{figure}
\centering\includegraphics[scale=0.9, trim= 1cm 15.5cm 1cm 0cm]{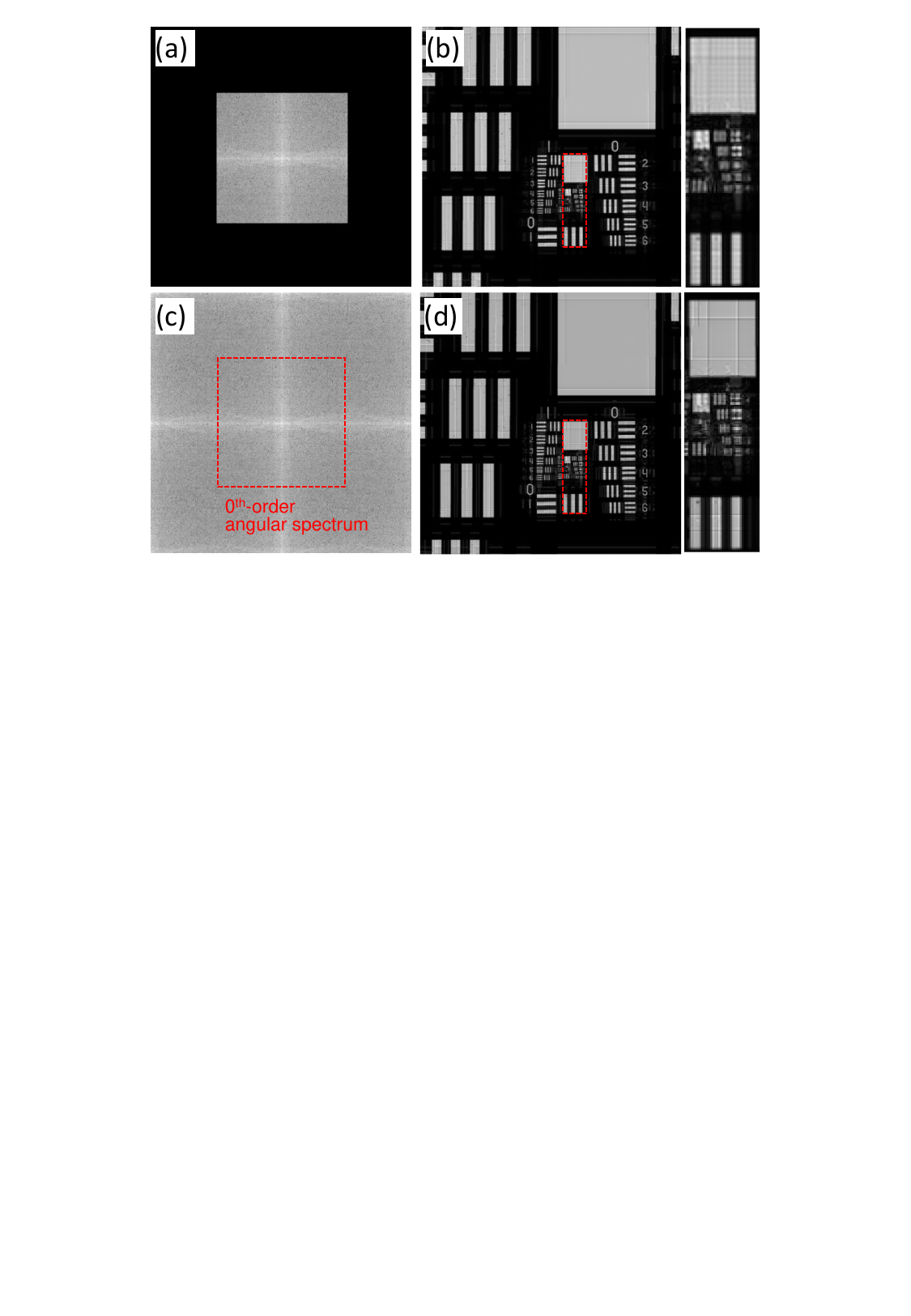}
\caption{Simulation results for image reconstruction of undersampled hologram. 
(a) Zero-padded angular spectrum into 512$\times$512 pixels in the logarithmic scale and (b) restored USAF target image through zero-padding technology. 
Right picture is the magnified image in red box. 
(c) 2$\times$2 expanded angular spectra by placing replica in an array. 
Red box denotes the zeroth order value. (d) Directly restored image via two-multiple expansion of angular spectrum in Fourier space.}
\end{figure}

\begin{figure}
\centering\includegraphics[scale=0.85, trim= 1cm 14cm 1cm 0cm]{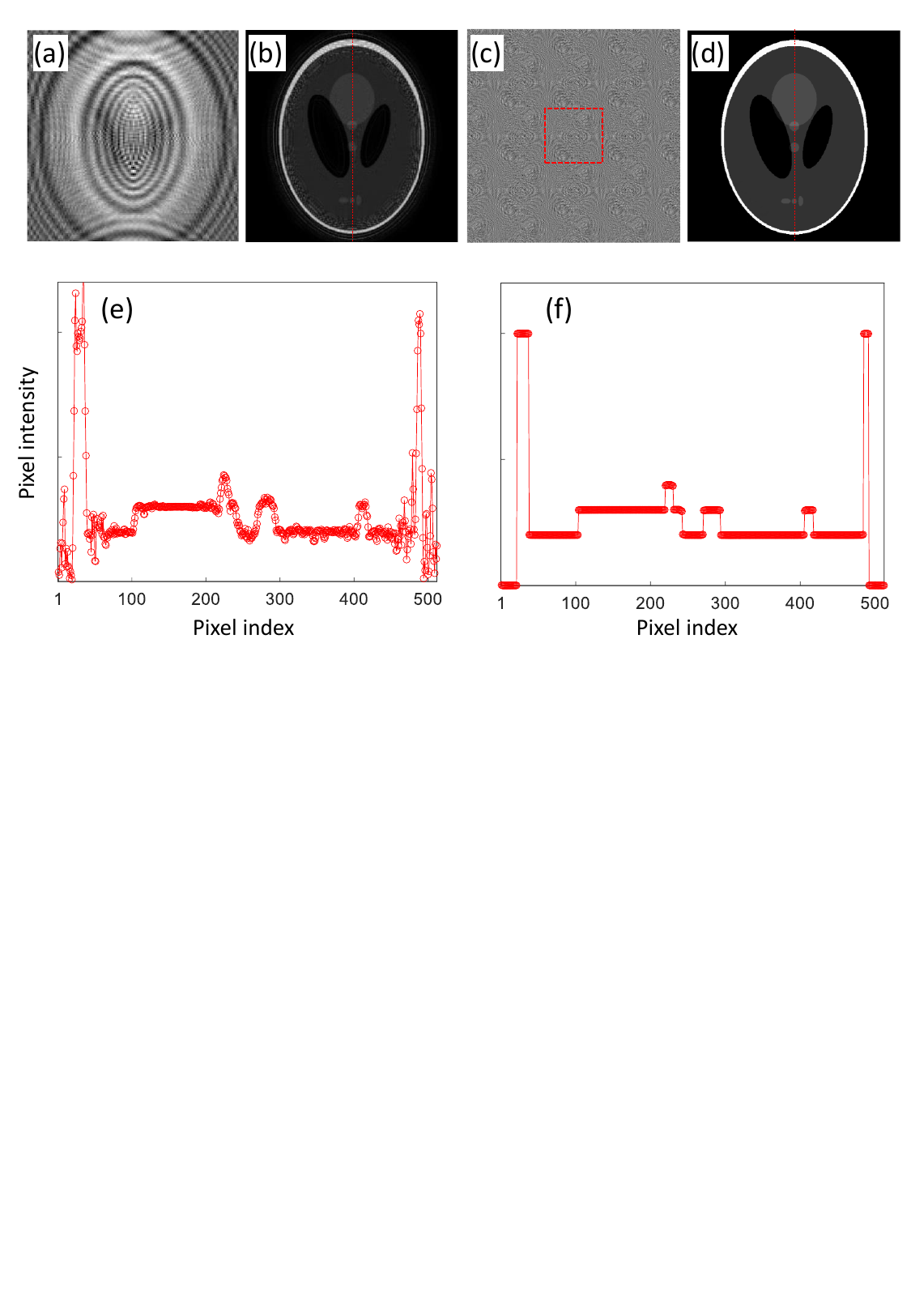}
\caption{Image reconstruction of undersampled hologram by using iterative opimization algorithm. 
(a) Low-resolution hologram with 128$\times$128 pixels of a 16-$\mu$m pixel interval for the Shepp-Logan phantom.
(b) Directly restored image via four-multiple expansion of angular spectrum.
(c) 4$\times$4 expanded angular spectra in the logarithmic scale after iterative optimization.
(d) Restored image via the iterative optimization algorithm including 4-fold expansion of angular spectrum.
Cross-sectional profiles of reconstructed images (e) without and (f) with iterative optimization algorithm.}
\end{figure}

\begin{figure}
\centering\includegraphics[scale=0.9, trim= 0cm 14.5cm 1cm 0cm]{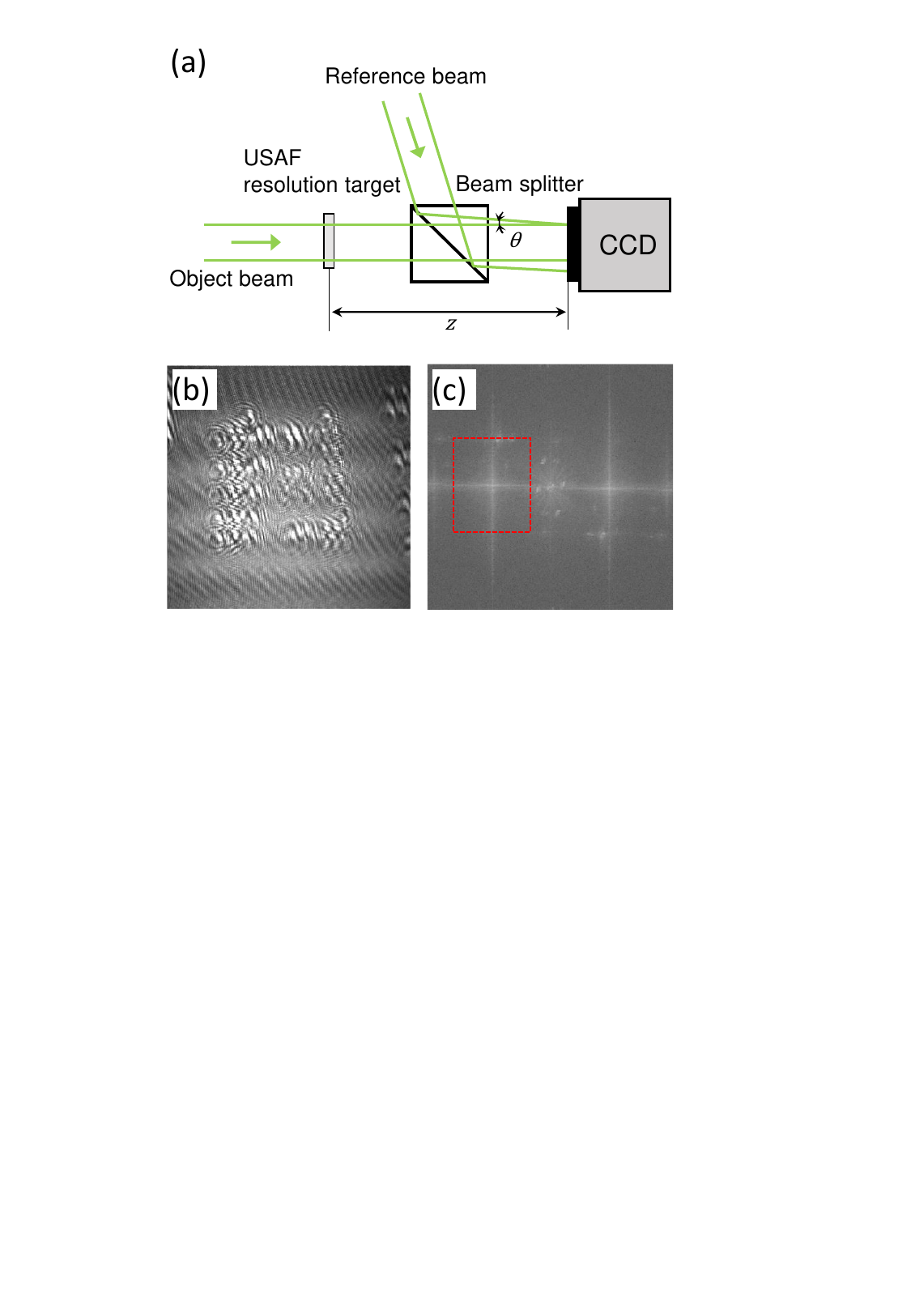}
\caption{(a) Off-axis holographic technique to acquire an optical hologram. (b) Captured hologram for the USAF resolution target
and (c) its Fourier-transformed data. Red box indicates the first-order region. $\theta$ is an off-axis angle.}
\end{figure}

\begin{figure}
\centering\includegraphics[scale=0.9, trim= 1cm 14cm 1cm 0cm]{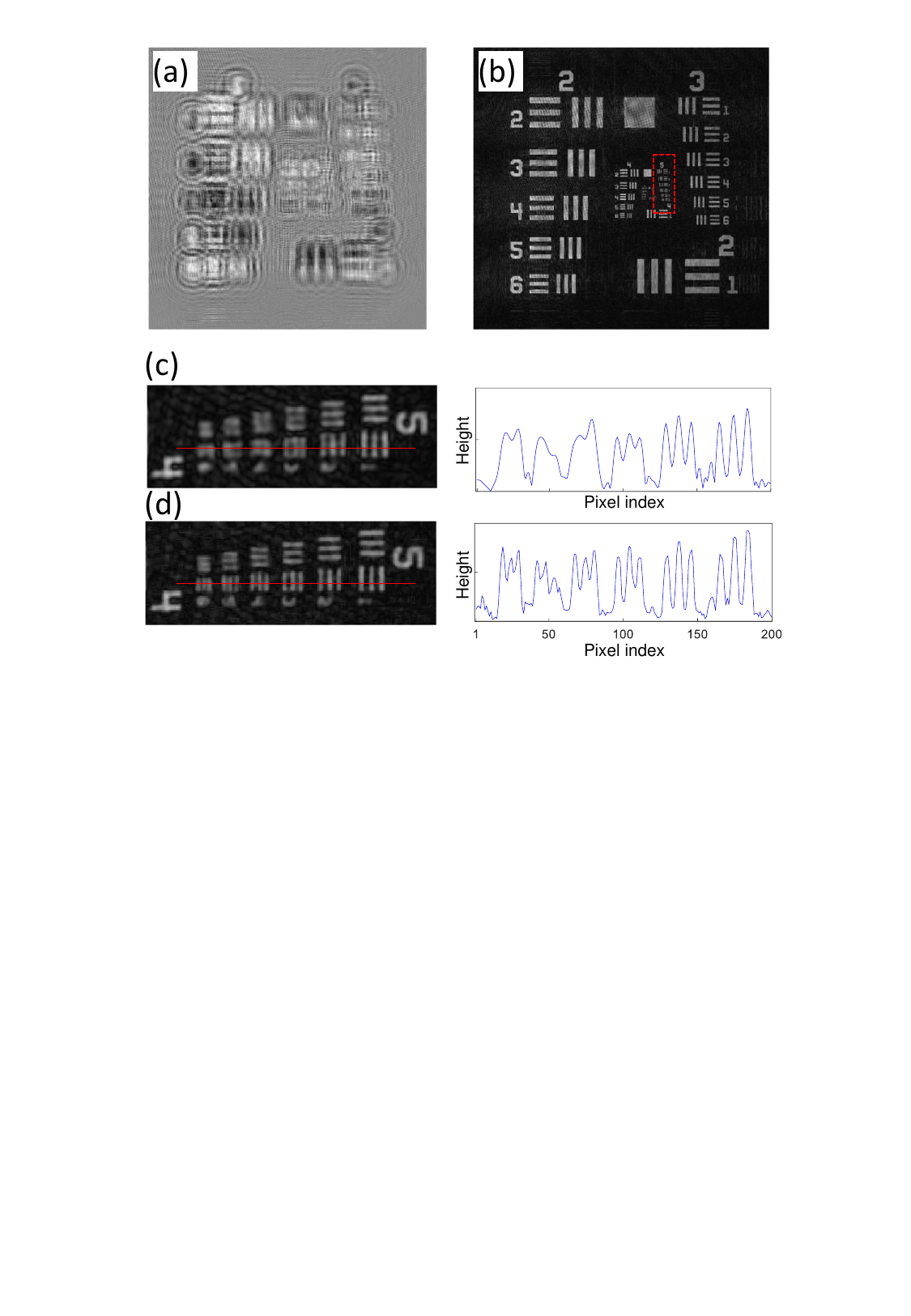}
\caption{Image reconstruction phenomenon of optical hologram. (a) Low-resolution optical hologram for USAF resolution target.
(b) Image restored via the iterative optimization algorithm including the expansion process of angular spectrum. 
Magnified images and profiles of pixel values for the restored images (c) without and (d) with the spectrum expansion process.}
\end{figure}

\end{document}